\documentclass{ieeeaccess}

\def\headerlogo{} \def\headerlogoall{} \headname{}
\usepackage{etoolbox} \makeatletter \patchcmd{\@maketitle} {{\vss\doifont Digital Object Identifier\space\@doi\vss}\vspace*{7.4mm}\par} {} {}{} \makeatother
\makeatletter \def\headerlogo{} \def\headerlogoall{} \def\ps@headings{%
 \def\@oddhead{\parbox[t]{\textwidth}{\mbox{}\\[-6.3mm] {\raisebox{-2pt}{\headerfont\rightmark}} \vspace*{-1.5mm}\par\hrulefill}} \def\@evenhead{\parbox[t]{\textwidth}{\mbox{}\\[-6.3mm] {\raisebox{-2pt}{\headerfont\leftmark}} \vspace*{-1.5mm}\par\hrulefill}} \def\@oddfoot{} \def\@evenfoot{} 
} \def\ps@titlepage{% 
\def\@oddhead{} \def\@evenhead{} \def\@oddfoot{} \def\@evenfoot{} 
} \makeatother

\makeatletter

\long\def\@makecaption#1#2{%
\ifx\@captype\@IEEEtablestring%
\begin{flushleft}
\vspace*{5pt}
{\vss\color{accessblue}\tablecapheadfont #1. \ }%
{\raggedright\tablecapfont#2\vss}%
\end{flushleft}
\@IEEEtablecaptionsepspace%
\else%
\@IEEEfigurecaptionsepspace%
\noindent%
\ifdim\xfigwd>\columnwidth
  \parbox{\textwidth}{%
    {\color{accessblue}\figcapheadfont #1. \ }%
    \figcapfont #2%
  }%
\else
  \parbox{\columnwidth}{%
    {\color{accessblue}\figcapheadfont #1. \ }%
    \figcapfont #2%
  }%
\fi%
\fi%
\vskip 1pt plus 1pt minus 1pt%
}

\makeatother

\usepackage{cite}
\usepackage{amsmath,amssymb,amsfonts}
\usepackage{algorithmic}
\usepackage{graphicx}
\usepackage{textcomp}
\usepackage{makecell}
\usepackage{hyperref}
\def\BibTeX{{\rm B\kern-.05em{\sc i\kern-.025em b}\kern-.08em
    T\kern-.1667em\lower.7ex\hbox{E}\kern-.125emX}}

\begin{document}

% this should be updated once the article is published!!!
%\history{Date of publication xxxx 00, 0000, date of current version xxxx 00, 0000.}
%\doi{10.1109/TQE.2020.DOI}  
\history{}
\doi{}

\title{
SpinControl: Control Software for Prototyping and Benchmarking of Spin‑Qubit Devices and Measurement Protocols
}
\author{
\uppercase{
Harold B. Meerwaldt\authorrefmark{1},
Daniël P. van der Velde\authorrefmark{1,2},
Pieter T. Eendebak\authorrefmark{1}, 
and Michiel J. Haye\authorrefmark{1},
}
}
\address[1]{Netherlands Organization for Applied Scientific Research (TNO), Delft, The Netherlands}
\address[2]{Embedded Customs, Zoetermeer, The Netherlands}

\markboth 
{Meerwaldt \headeretal: SpinControl}
{Meerwaldt \headeretal: SpinControl}

\corresp{Corresponding author: Pieter T. Eendebak (email: pieter.eendebak@tno.nl).}
\tfootnote{Author manuscript version, 16 October 2026}

\begin{abstract}
Quantum computers require software to control the states of the qubits.
As the actions on qubits grow in complexity with circuit execution,
calibration, and benchmarking, the complexity of the control software
grows in tandem. Spin-qubit quantum computers are at a stage where
experimentalists need low-level access for tuning and calibration, and
quantum algorithm researchers want high-level circuit-level access. In
this article, we present the software design of SpinControl. SpinControl
integrates three architectural metrics: 1) \textit{customizability,} to
allow prototyping devices and measurement protocols, is achieved by
separating data and behaviour; 2) \textit{modularity} and \textit{extensibility}
are achieved by separating functionalities into different components;
and 3) \textit{robustness,} to reliably execute quantum circuits on many
qubits, is achieved by CI pipelines and input validation. SpinControl
allows different types of measurements while customizing the instruments
used, the pulse or circuit program executed, and the data processing. It
allows building standardized experiments from measurement and analysis
building blocks and executing these for calibration in a custom and
conditional order. SpinControl implements these functionalities
alongside robust configuration and data management, a connection to the
cloud, and calibration scheduling. Pulse and circuit compilation,
virtual gate matrices, as well as spin-qubit-specific GUIs integrated by
SpinControl, make heavy use of software packages developed by Delft
University of Technology: pulse-lib, core-tools (both open source), and
QConstruct.
\end{abstract}

\begin{keywords}
control software, quantum computing, software architecture, spin qubits
\end{keywords}

\titlepgskip=-15pt

\maketitle

\makeatletter \def\ps@myplain{%
 \def\@oddhead{}% 
\def\@evenhead{}% 
\def\@oddfoot{\hfill\thepage\hfill}% 
\def\@evenfoot{\hfill\thepage\hfill}% 
} \makeatother

\pagestyle{myplain}

\section{Introduction}
\label{sec:introduction}

Quantum computers have the potential to perform calculations that are
intractable on classical computers \cite{Childs10,Mon16,Nie10}.
Semiconductor spin qubits \cite{Van19,Bur23,DeM23} are considered a
promising platform to reach the many quantum bits (qubits) needed to
perform these calculations \cite{Blu19}, because of their long coherence
times \cite{Sta22}, small size, and compatibility with CMOS processes
\cite{Stu24}. Operating a semiconductor spin-qubit system requires tuning
\cite{Sch26}, calibrating \cite{Phi22}, benchmarking \cite{Sta22}, and
executing algorithms \cite{Phi22,Hua24}. These diverse activities
require software that offers large customizability of operation and
deals with different abstractions in a maintainable manner.

Semiconductor spin qubits are commonly controlled with
nanosecond-resolution voltage pulses and acquisitions
\cite{Vel14,Kaw14}. The control electronics that perform these pulses
and acquisitions depend on different hardware technologies. The
electronics can be CMOS-based, with the aim of cryogenic operation
\cite{Gec20,Pra22}. The electronics can be based on FPGAs
\cite{Tou24,Fru24} or RFSoCs \cite{Din24,Par22,XuY23} running on
firmware \cite{Geb21}, with the aim of low-latency readout. As a third
option, off-the-shelf high-sample rate instruments like arbitrary wave
generators and lock-in amplifiers are used. Specific to semiconductor
spin qubits, control also requires ultra-stable DC voltages, besides
rapidly varying voltage pulses.

In turn, experiment control software is used to control and read out the
control electronics (and other instruments) at a higher abstraction
level. The different levels of abstraction lead to a layered
architecture for quantum computers as a whole \cite{Jon12}. The
experiment control software follows this layered architecture
\cite{Guo19} or is modular \cite{Rie22}, and can extend control from
the device up to a cloud service \cite{Ahm24}. Experiment control
software is part of a large body of quantum software \cite{Ser22}, with
layered and pipe-and-filter as the most common software architectures
\cite{Kha23}.

In this article, we introduce SpinControl, an experiment control
software package specifically focused on the operation of spin-qubit
systems. Written in Python, it acts as a library for maintainers of the
system to tune it to a working state, calibrate it to keep it in a
working state, and benchmark it to assess the quality of the working
state. The library is used by experimentalist researchers to prototype
new spin-qubit devices and measurement protocols. At the same time,
SpinControl acts as a component in a larger system, which allows
external users to send quantum algorithms through a cloud platform and
receive the results back. Fig.~\ref{fig1} gives a high-level overview of the operation
of a spin-qubit system with both internal users (maintainers and
experimentalists) and external users.

\Figure[t!](topskip=0pt, botskip=0pt, midskip=0pt)[width=6.12205in]{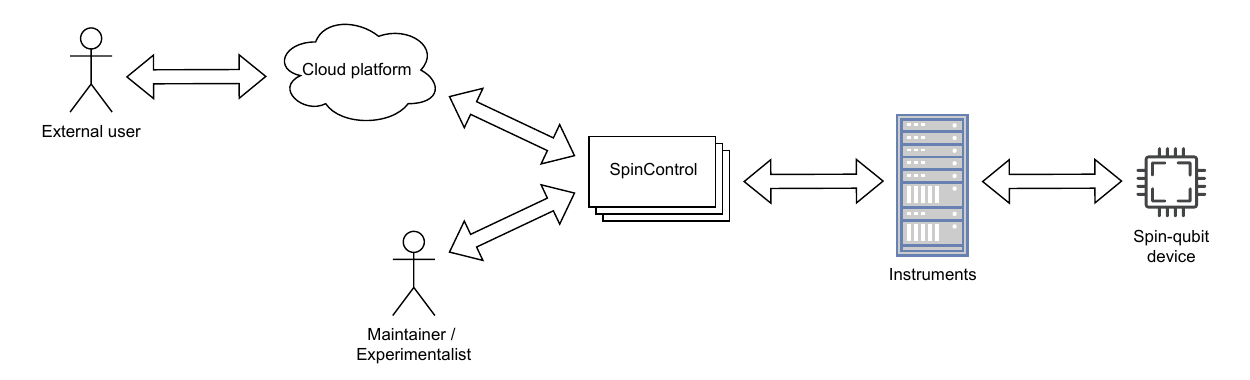}
{\label{fig1} High-level overview of operation of a spin-qubit system using
SpinControl}

The dual aim of being a library for maintainers and researchers and
being a component in a larger system has led to SpinControl meeting
several software architecture metrics:

\begin{itemize}
\item
  High \textbf{customizability} and \textbf{traceability} are achieved
  by having a generic core and having the user pass functionality to the
  core, thus separating the concerns of data and behaviour. To be
  described in more detail in this article, the user can supply device
  layout, instruments (of different manufacturers) with parameters,
  pulse or circuit programs, pre-storage processing of results, and
  more. Another means of customizing is by combining elements in the
  form of standardized experiments and calibration graphs.
\item
  A \textbf{modular} and \textbf{extensible} architecture with low
  coupling and high cohesion is achieved by separating functionalities
  into different components (see Section~\ref{components}).
\item
  \textbf{Robustness} is achieved in continuous integration (CI)
  pipelines and during runtime. The CI pipelines verify the correct
  execution of the software with unit, integration, and system tests,
  type-checking, and linting. During runtime, input validation gives
  feedback before failure can occur. Errors during calibration are
  suppressed and logged, allowing execution to continue.
\end{itemize}

Besides software robustness, robustness of the spin-qubit system is
achieved through automated calibration, which repairs the system after
an out-of-spec calibration. Measurement reproducibility is achieved in
two ways. Firstly, configuration of the system and user scripts are
under version control. Secondly, measurement results are stored together
with input measurement variables, a snapshot of instrument and other
system settings, and possible analysis on the measurement results.

Besides SpinControl, other experiment control software exists that can
be used for spin-qubit systems, summarized in Table \ref{tab:software-comparison}.

\begin{itemize}
\item
  Q-CTRL's Boulder Opal \cite{QCT26} and QuantrolOx' EDGE
  \cite{Qua26} provide pulse and circuit compilation, and automated
  tuning and calibration using standardized experiments. These
  functionalities are not implemented for spin qubits specifically.
\item
  Qblox' qblox-scheduler \cite{Qbl}, Zurich Instruments' LabOne Q
  \cite{Zur26}, and Quantum Machines' QUA \cite{Qua261}
  provide pulse compilation for their own control electronics. Some
  include circuit compilation, non-spin-specific automated calibration
  and standardized experiments or circuits, or a framework for
  automated tuning.
\item
  spinQICK \cite{HRL} has pulse compilation and standardized
  experiments focussed on spin-qubits. It only supports one specific
  brand of RFSoC-based hardware.
\item
  Orange QS' Juice \cite{Ora26} combined with Quantify
  \cite{Ora} is hardware-agnostic and has pulse and circuit
  compilation. It is not spin-specific.
\item
  Delft University of Technology's stack of core-tools
  \cite{Del26}, pulse-lib \cite{Del261}, and QConstruct
  (closed source) has support for virtual gate matrices, is
  hardware-agnostic, does pulse and circuit compilation, and contains
  non-spin-specific standardized circuits.
\end{itemize}

\begin{table*}[t] \caption{Comparison of control software packages for quantum computers.
``vendor-only'': only electronics of the vendor can be controlled with
the software. ``not spin'': experiments, calibration, or tuning is
present but not specifically for spin qubits. ``non-spin circuits'': the
experiments have circuits implemented that are not specific for spin
qubits. ``framework'': the software contains a framework but no
implementation for automated calibration or tuning.} 
\label{tab:software-comparison} 
\centering
\small
\setlength{\tabcolsep}{4pt}
\renewcommand{\arraystretch}{2}
\resizebox{\textwidth}{!}{% 
\begin{tabular}{|l|c|c|c|c|c|c|c|c|c|c|} \hline
 & \multicolumn{3}{c|}{Instrumentation} & \multicolumn{3}{c|}{\makecell[c]{Compilation\\and execution}} & \multicolumn{3}{c|}{Calibration} & Cloud \\ \cline{2-11}

 & \rotatebox[origin=c]{90}{Control electronics} & \rotatebox[origin=c]{90}{Virtual gate matrices} & \rotatebox[origin=c]{90}{  Device gate-channel mapping  } & \rotatebox[origin=c]{90}{  Pulse generation/compilation  } & \rotatebox[origin=c]{90}{Circuit compilation} & \rotatebox[origin=c]{90}{Pre-storage processing} & \rotatebox[origin=c]{90}{Standardized experiments} & \rotatebox[origin=c]{90}{Automated calibration} & \rotatebox[origin=c]{90}{Automated tuning} & \rotatebox[origin=c]{90}{Cloud algorithms} \\  \hline

\makecell[l]{SpinControl\\(this article)} & \makecell[c]{$\checkmark$\\\ref{instruments}} & \makecell[c]{$\checkmark$\\\ref{instruments}} & \makecell[c]{$\checkmark$\\\ref{instruments}} & \makecell[c]{$\checkmark$\\\ref{compilation}} & \makecell[c]{$\checkmark$\\\ref{compilation}} & \makecell[c]{$\checkmark$\\\ref{execution}} & \makecell[c]{$\checkmark$\\\ref{experiments-and-calibration}} & \makecell[c]{$\checkmark$\\\ref{experiments-and-calibration}} & \makecell[c]{framework\\\ref{experiments-and-calibration}} & \makecell[c]{$\checkmark$\\\ref{orchestration}} \\ \hline

Boulder Opal & $\checkmark$ & & $\checkmark$ & $\checkmark$ & $\checkmark$ & $\checkmark$ & not spin & not spin & not spin & $\checkmark$ \\ \hline 

EDGE & $\checkmark$ & & $\checkmark$ & $\checkmark$ & $\checkmark$ & $\checkmark$ & not spin & not spin & not spin & $\checkmark$ \\ \hline qblox-scheduler & vendor-only & & & $\checkmark$ & $\checkmark$ & & non-spin circuits & & & \\ \hline 

LabOne Q & vendor-only & & & $\checkmark$ & $\checkmark$ & $\checkmark$ & not spin & not spin & frame-work & \\ \hline QUA & vendor-only & $\checkmark$ & $\checkmark$ & $\checkmark$ & & & & not spin & & \\ \hline SpinQICK & single RFSoC & & $\checkmark$ & $\checkmark$ & & & $\checkmark$ & & & \\ \hline 

Juice + Quantify & $\checkmark$ & & & $\checkmark$ & $\checkmark$ & & non-spin circuits & frame-work & & $\checkmark$ \\ \hline 

\makecell[l]{core-tools +\\ pulse-lib +\\ QConstruct} & $\checkmark$ & $\checkmark$ & $\checkmark$ & $\checkmark$ & $\checkmark$ & & non-spin circuits & & & $\checkmark$ \\ \hline 
\end{tabular}%
}
\end{table*} 

Adjacent to experiment control software, Qiskit \cite{IBM26},
PennyLane \cite{Vil18}, and OpenSquirrel \cite{Del262} are frameworks
for constructing and compiling quantum algorithms. Qiskit, PennyLane,
Conductor Quantum \cite{Con26}, Classiq \cite{Cla26}, and qBraid
\cite{qBr26} allow executing algorithms through multiple cloud
platforms. None of these allow control of electronics directly.

\section{High-level architecture}\label{high-level-architecture}

To improve cohesion and have clear dependencies, SpinControl is
structured in a modular fashion, both between and within Python
packages. SpinControl is divided into the following packages, with their dependencies as shown in Fig.~\ref{fig2}.

\begin{itemize}
\item
  The \textit{low-level-software} package holds functionality closest to
  the instruments, including instrument drivers, compilation,
  measurement execution, and more, as will be explained in more detail
  in Section~\ref{components}.
\item
  The \textit{analysis} package holds different types of analyses that are
  performed on the outcomes of measurements.
\item
  The \textit{lls-experiments} package (for \textit{low-level-software}
  experiments) holds standardized experiments, combining functionality
  from \textit{low-level-software} and \textit{analysis}.
\item
  The \textit{llscfg} package (for \textit{low-level-software}
  configuration) holds the configuration files for each device/hardware
  stack and the scripts used by experimentalists for tuning and
  benchmarking, making use of the above three packages.
\item
  The \textit{utils} package holds utility functionality used by all other
  packages.
\end{itemize}

\Figure[t!](topskip=0pt, botskip=0pt, midskip=0pt)[width=3.37402in]{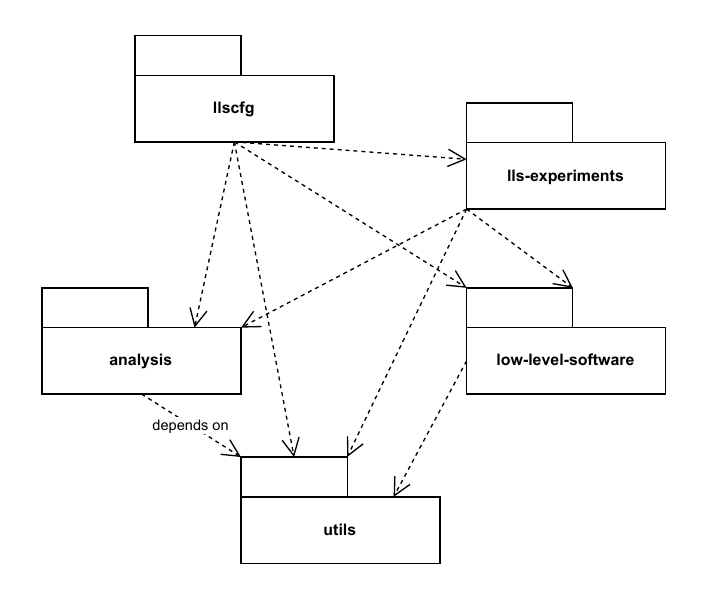}
{\label{fig2} Dependencies between Python packages in SpinControl}

The \textit{low-level-software} package separates the following
functionalities into components. The dependencies between the components
are shown in Fig.~\ref{fig3}.

\begin{itemize}
\item
  \textbf{Instrumentation}. The InstrumentManager component is the
  access point to the instruments through the instrument drivers.
\item
  \textbf{Compilation}. The Compiler component converts gate-based
  circuits and pulse schedules before uploading them to instruments and
  retrieves the results.
\item
  \textbf{Execution}. The MeasurementManager component allows different
  types of measurements such as scans and executing pulse schedules,
  combining Instrumentation and Compilation.
\item
  \textbf{Calibration}. The CalibrationManager component schedules and
  orders the execution of experiments for calibration of the device
  settings. Many experiments are standardized.
\item
  \textbf{Cloud connection}. The OnlineManager component maintains the
  connection to a cloud platform to receive quantum algorithms from
  external users and return results.
\item
  \textbf{Configuration, data management, and graphical user interfaces
  (GUIs)}. The StorageManager component retrieves the configuration of
  other components through configuration files. The output data of the
  execution is stored to disk. Internal users can modify settings and
  see outputs with the use of several GUIs.
\item
  \textbf{Orchestration}. The LlsMain class initializes and configures
  the components above.
\end{itemize}

\Figure[t!](topskip=0pt, botskip=0pt, midskip=0pt)[width=4.07874in]{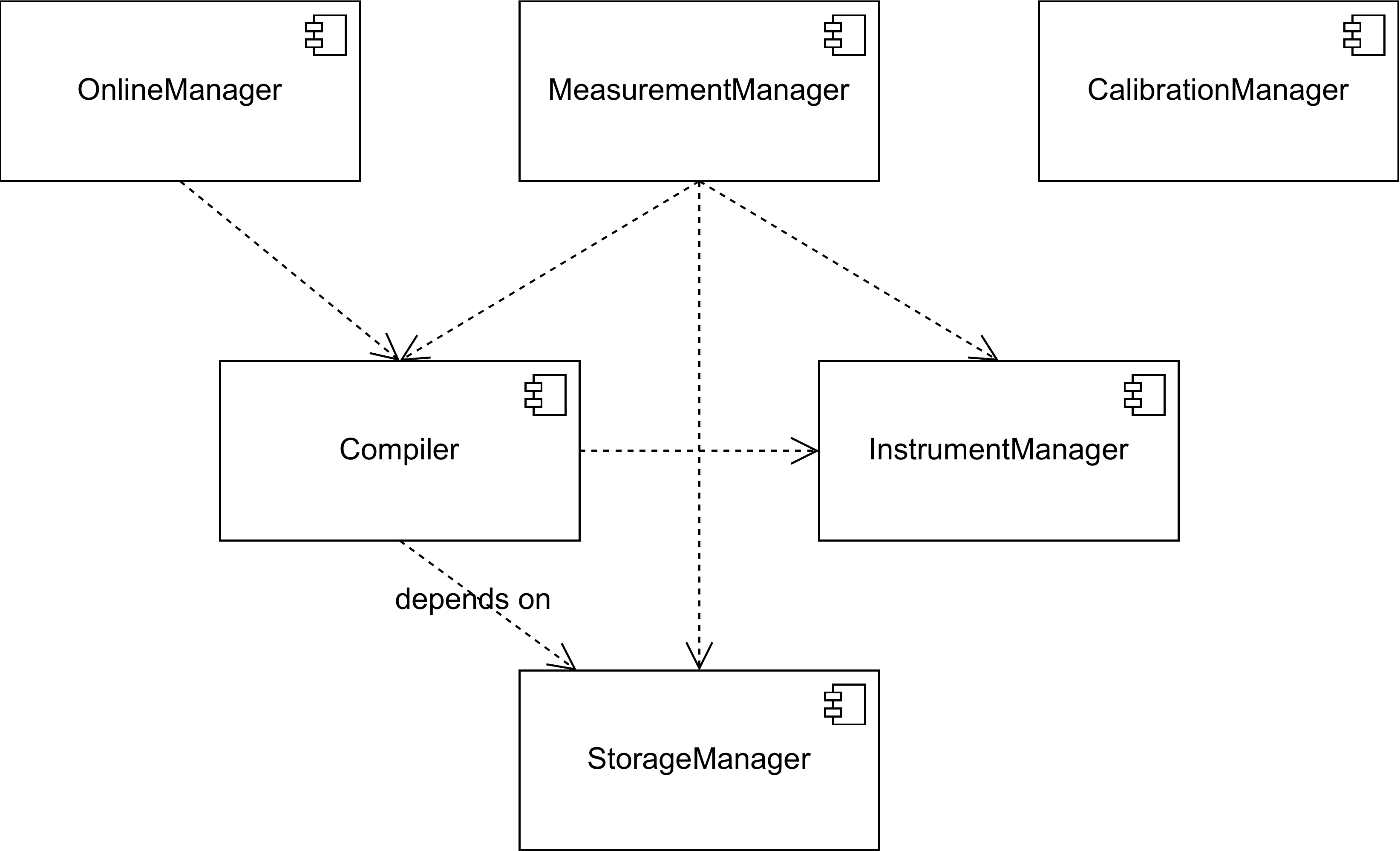}
{\label{fig3} Dependencies between components in SpinControl}

\section{Components}\label{components}

\subsection{Instruments}\label{instruments}

Instruments (also known as hardware or electronics) and their settings
are represented in software by instrument drivers. Instrument drivers
hold the communication protocol and additional logic to control and read
out the state of the instrument. The \texttt{InstrumentManager} class loads the
configuration file for the instrument drivers and instantiates them.

The instrument drivers are based on the \textit{qcodes} package \cite{Mic26}.
\textit{qcodes} provides an \texttt{Instrument} class where you can add
\texttt{Parameter} objects with which you can set or get the value of an
instrument setting.

Instrument drivers are kept within the \textit{low-level-software}
package, as subclasses of implementations from external packages such as
\textit{qcodes} itself or \textit{qcodes\_contrib\_drivers}. Some instrument
drivers are expanded considerably with custom functionality.

All implemented instrument drivers have a simulator counterpart, to
allow the instruments to be used in system tests and documentation
notebooks as fakes. Some simulators have pre-recorded return values for
getting and ignore setting; some simulators use \textit{qcodes'} pyvisa
simulation file; some simulators allow setting the returned simulated
data by the user.

A \texttt{PortInstrument} is a virtual (i.e. non-physical)
\texttt{Instrument} that maps a port (also known as gate or terminal) on
a device to a channel on an instrument. This allows the user to use
functional port naming in experiments instead of physical instrument
channels, hereby reducing errors. The user can modify this mapping in a
configuration file.

There are other helper instruments that group parameters that do not
belong to a physical instrument. The \texttt{AuxInstrument} holds
parameters to record the time or get a value from a simple counter. The
\texttt{UserparamInstrument} acts as a container for users to register
their own parameters to.

The \texttt{InstrumentManager} has a \texttt{VPortManager} that handles
virtual gate matrices supplied by the \textit{core-tools} \cite{Del26} and
\textit{pulse-lib} \cite{Del261} packages. Virtual gate matrices are a
functionality specifically needed to operate semiconductor spin qubits,
compared to other quantum-computing modalities. Through the matrix,
setting a voltage on a virtual port results in voltages on multiple real
ports. The virtual gate matrices are stored in a configuration file,
which can be modified through a graphical user interface.

The relations between these classes are shown in Fig.~\ref{fig4}.

\Figure[t!](topskip=0pt, botskip=0pt, midskip=0pt)[width=5.35433in]
{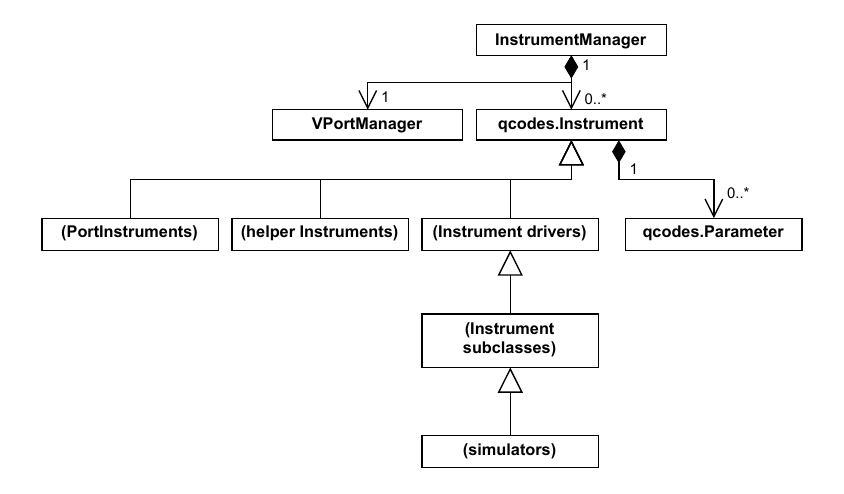}
{\label{fig4} Class diagram of InstrumentManager, Instrument, and related classes}

\subsection{Compilation}\label{compilation}

The user can create programs to be run on arbitrary waveform generators
and high-frequency lock-in amplifiers. The programs set and read out
voltages with up to nanosecond resolution. These steps are performed by
the \texttt{Compiler}, which consists of a \texttt{PulseCompiler} and a
\texttt{CircuitCompiler}. Hardware backends from multiple manufacturers
are supported.

The \texttt{PulseCompiler} provides a program for the user at the pulse
level, specifying voltages and acquisitions directly. The user can
modify the pulse program by adding different segments like block pulses,
ramps, custom pulses, and acquisitions. The \texttt{PulseCompiler}
handles all configuration settings for compilation (like averaging and
sample rate), converts the pulse program to a format that the specific
hardware backend understands, and uploads it. After running the pulse
program, the hardware sends the results from acquisition back to the
\texttt{PulseCompiler}. The \texttt{PulseCompiler} is a wrapper around
the \textit{pulse-lib} package \cite{Del261} and can configure
\textit{pulse-lib} to compensate for charging effects in the electronic
circuit that combines the DC and RF signals to the device gates.

One specific feature is parametrization, where a sequence of values
(e.g. voltage or duration) is used instead of a single value. The
\texttt{PulseCompiler} then converts the pulse program into a sequence
of programs, changing the parametrized value for each program. If
multiple values are parametrized, the sequence of programs contains all
combinations of values.

Besides handling pulse programs, the \texttt{PulseCompiler} also holds
settings, like the lock-in amplifier's output amplitude and LO
frequency. The pulse-lib hardware backends are not \textit{qcodes}
\texttt{Instrument} subclasses, and to vary the settings in experiments,
\textit{qcodes} \texttt{Parameter} objects are needed. The
\texttt{PulseCompilerInstrument} is a bridging \textit{qcodes} instrument
driver that holds \texttt{Parameter} objects that map to
\texttt{PulseCompiler} settings.

The \texttt{CircuitCompiler} provides a program for the user at the
circuit level, specifying quantum gates or other instructions. Each
instruction has an implementation at pulse level, which can be
different for each qubit. Some instructions (like Rz) have a variable
that can be modified when constructing a circuit program, and
parametrization is also possible. The circuit program is converted into
a pulse program for the \texttt{PulseCompiler} to handle. The
\texttt{CircuitCompiler} is a wrapper around the closed-source
\textit{QConstruct} package, developed by Delft University of Technology.

Many circuit-level instructions have settings, e.g. the duration,
voltage, and frequency of an \texttt{X90} gate. These settings are
stored in a datastore. A copy of the datastore is stored with each
experiment, so the exact settings can be used to reproduce the
experiment.

The \texttt{CircuitCompiler} supports the cQASM quantum assembly
language \cite{QuT26} that is passed down from the Quantum Inspire
platform \cite{QuT261}. The cQASM string is converted into a circuit
program consisting of \textit{QConstruct} instructions. SpinControl
shares the primitive gates (i.e. the gates that the spin-qubit system
can execute without decomposition) with the Quantum Inspire platform at
configuration time, which in turn ensures that the cQASM string only
contains these gates.

The \texttt{DeviceManager} handles the configuration of the device,
including available qubits and qubit pairs (connectivity). It also
registers the implementations (per qubit or qubit pair) of the
circuit-level instructions and the settings belonging to each
instruction in the settings datastore. SpinControl contains
implementations (per qubit) for initialization and readout of
arbitrary-length linear arrays of spin-qubits. The
\texttt{DeviceManager} uses a plug-in architecture, so the user can
provide a device configuration tailored to the device. The
\texttt{DeviceManager} is used by the \texttt{CircuitCompiler} to
convert a circuit-level program into a pulse program.

Fig.~\ref{fig5} shows the steps to execute a pulse program, circuit
program, or cQASM string, and receive the acquisitions.

\Figure[t!](topskip=0pt, botskip=0pt, midskip=0pt)[width=5.90551in]{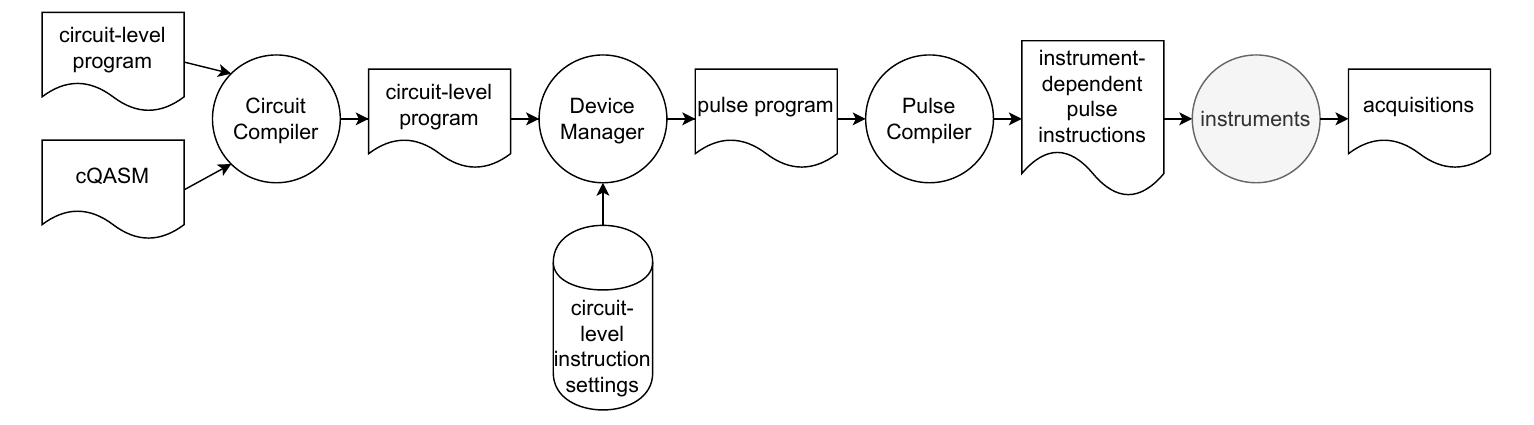}
{\label{fig5} Data flow diagram from circuit-level or cQASM program to acquisitions. A
circuit-level program or cQASM string is passed to the CircuitCompiler,
which passes the circuit-level program untouched or converts the cQASM
string into a circuit-level program. The DeviceManager converts the
circuit-level program into a pulse program using settings from a
datastore. The PulseCompiler converts the pulse program into
instrument-dependent pulse instructions, which are executed on the
instruments, returning acquisitions.}

\subsection{Execution}\label{execution}

When performing measurements, hardware settings are varied and hardware results are recorded. The \texttt{Measurement\-Manager} class is a façade to perform different types of measurements, as shown in Fig.~\ref{fig6}a.

\begin{itemize}
\item
  \textbf{Scan}. One or multiple (\textit{qcodes})
  parameters are varied in one or more dimensions. For each
  combination of parameters set, the value of one or multiple
  parameters is retrieved.
\item
  \textbf{AwgScan}. The arbitrary waveform generator is programmed with
  a pulse program containing a sawtooth for one or two instrument
  channels, while a sequence of return values is recorded in a batch. To
  minimize electronic circuit charging effects, a pulse program with an
  alternating staircase can also be programmed.
\item
  \textbf{PulseMeasurement}. A pulse program is sent to the
  \texttt{Compiler} alongside other settings and uploaded to the
  hardware after which the results are recorded in a batch.
\item
  \textbf{CircuitMeasurement}. A circuit-level program is sent to the
  \texttt{Compiler} alongside other settings and uploaded to the
  hardware after which the results are recorded in a batch.
\item
  \textbf{QubitMeasurement}. A circuit-level program is sent to the
  \texttt{Compiler}. The results are processed to give single-qubit
  states, fractions (relative frequency) for single qubits or
  multi-qubit bitstrings, or custom processing is performed.
\item
  \textbf{CqasmMeasurement}. A cQASM string is sent to the
  \texttt{Compiler}. The results are processed to give multi-qubit
  bitstrings and fractions.
\end{itemize}

The last four measurements build on each other, forming layers based on
the level of abstraction \cite{Jon12}: a \texttt{CqasmMeasurement}
forwards to a \texttt{Qubit\-Measurement}, which forwards to a
\texttt{Circuit\-Measurement}, which forwards to a
\texttt{Pulse\-Measure\-ment}. These four measurements all result in
executing a pulse program. Both settings of the pulse program and other
settings can be parametrized (looped).

Execution in SpinControl allows high customizability by separating data
and operation. This is achieved by passing input definitions to the
\texttt{MeasurementManager}'s façade methods. The
\texttt{MeasurementBuilder} uses the values of the input definitions,
such as parameter names, value ranges, or pulse/circuit programs to
build \texttt{Parameter} objects that can be set and gotten by a generic
\texttt{Measurement} class. These \texttt{Parameter} objects are
retrieved from the \texttt{InstrumentManager} by parameter name or
constructed to set and get from a pulse program.

\Figure[t!](topskip=0pt, botskip=0pt, midskip=0pt)[width=4.68898in]{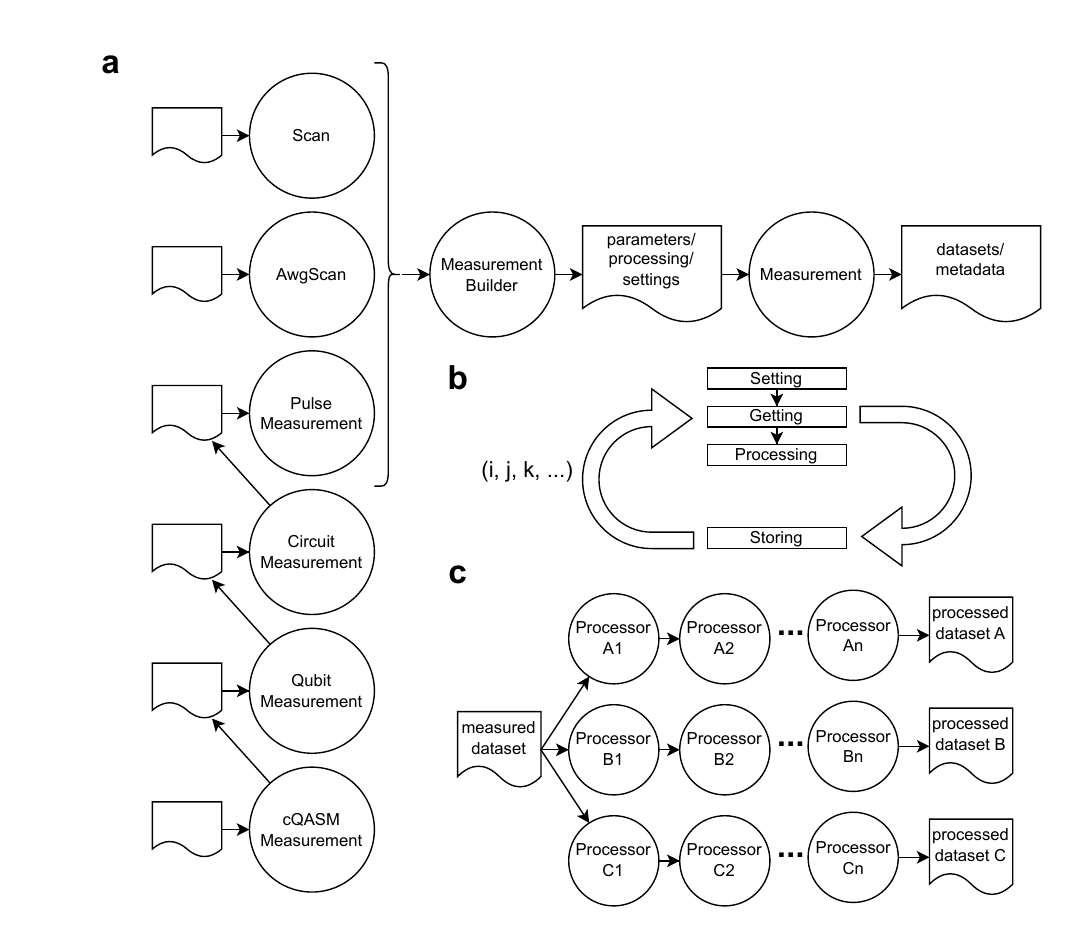}
{\label{fig6} Representation of execution, with a) data flow diagram, showing inputs
to different measurement types being converted by the MeasurementBuilder
to parameters, processing, and settings, before being used by a
Measurement in the creation of datasets, b) iteration of setting,
getting, processing, and storing for each measurement point, and c)
processing by multiple processor sequences from measured dataset to
processed datasets.}

The \texttt{Measurement} class contains a multi-dimensional loop with
three steps per iteration, as shown in Fig.~\ref{fig6}b.

\begin{enumerate}
\def\labelenumi{\arabic{enumi}.}
\item
  \textbf{Setting}. One or more \texttt{Parameter} values are set. If
  the \texttt{Parameter} value would be unchanged, this can be skipped
  if chosen by the user. \texttt{Parameter} objects can have their
  values vary in different dimensions or in the same dimension.
\item
  \textbf{Getting.} \texttt{Parameter} values are retrieved.
\item
  \textbf{Processing.} The retrieved \texttt{Parameter} values are
  processed.
\end{enumerate}

The \texttt{Measurement} class uses the \texttt{Dataset} from the
\textit{xarray} package \cite{Hoy17} as the data format to store values
in. After each user-settable update interval, the \texttt{Dataset} is
written to disk. Writing to disk during the measurement allows live
plotting and prevents losing data in case the measurement is aborted or
crashes.

The user can use \texttt{Processor} objects to customize
\textit{pre-storage} processing that is performed on the recorded values
before storing to disk. \texttt{Processor} objects perform operations on
the data and metadata of the recorded \textit{xarray}
\texttt{Dataset}. These operations can involve changing the number of
dimensions of the dataset, e.g. taking the average or performing a fit.
The \texttt{Processor} classes are kept small and modular, making them
more reusable. The \texttt{Processor} base class implements the Template
Method pattern, where the user only overrides those methods needed and
keeps common functionality.

The user provides \texttt{Flow} objects which are sequences of
\texttt{Processor} objects, so that the recorded \texttt{Dataset}
undergoes multiple transformations in sequence. As shown in Fig.~\ref{fig6}c, the user
can provide multiple \texttt{Flow} objects, which results in multiple
\texttt{Dataset} objects from the \texttt{Measurement} to be stored,
each with different processing applied.

To allow live plotting and intermittent storage, processing is attempted
at every iteration. To prevent processing on a dimension of data that
has not finished recording, a \texttt{Flow} has a trigger dimension. For
example, taking the mean of a trace has no meaning before the trace is
finished, and is not executed in that case.

\texttt{MeasurementStrategy} objects are used to combine the creation or
modification of a circuit with accompanying processing, since processing
is often strongly dependent on circuit specifics. The resulting circuit
and processing are then passed to \texttt{CircuitMeasurement} or
\texttt{QubitMeasurement}. \texttt{MeasurementStrategy} subclasses have
been implemented for parity-aligned readout, tomography readout, and
dynamic thresholding (where the voltage readout threshold between the 0
and 1 state varies during the measurement).

\subsection{Experiments and
calibration}\label{experiments-and-calibration}

Measurement, analysis, and more can be combined in an instance
of the \texttt{LlsExperiment} class. The \texttt{LlsExperiment} class
has template methods to configure the experiment, measure, analyse the
results, and generate plots. The class can be extended to update the
datastore and perform calibrations, checks (small measurements instead
of full calibrations), or benchmarks.

Standardized experiments based on \texttt{LlsExperiment} reside in the
\textit{lls-experiments} package and include the experiments in Table~\ref{tab:standardized-experiments}. The
measurements in these experiments are performed through one of the
measurement types of the \texttt{MeasurementManager}. The analyses used
in these experiments are located in the \textit{analysis} package.

\begin{table*}[t]
\caption{Standardized experiments in \emph{lls-experiments} and \emph{llscfg}.}
\label{tab:standardized-experiments}
\centering
\small
\begin{tabular}{ >{\raggedright\arraybackslash}p{0.24\textwidth} >{\raggedright\arraybackslash}p{0.16\textwidth} >{\raggedright\arraybackslash}p{0.24\textwidth} >{\raggedright\arraybackslash}p{0.30\textwidth} }
\hline
\textbf{Tuning} &
\textbf{Hardware} &
\textbf{Calibrations} &
\textbf{Benchmarks} \\
\hline

\makecell[tl]{
Charge sensor SNR\\
Chirp sweep\\
DGauss\\
RF resonance\\
Sensing dot\\
Signal properties\\
Gate response\\
Gate ratio\\
VGM tuning\\
Coulomb peak optimizer\\
Addition/polarization line scan
}
&
\makecell[tl]{
Fridge temperature\\
Pulsed RTS\\
Latency\\
IQ
}
&
\makecell[tl]{
CROT\\
Exchange strength and symmetry\\
CPhase/W angle and crosstalk\\
J vs B\\
Joff\\
Parallel Larmor\\
X90 phase\\
Rabi (frequency/amplitude/time/2d)\\
Shuttle
}
&
\makecell[tl]{
AllXY\\
Tomography (quantum state/gate set/\\
\hspace*{1em}quantum process/bilinear parity readout/\\
\hspace*{1em}3-qubit parity readout/parity-based)\\
Randomized Benchmarking (single-qubit/\\
\hspace*{1em}character/interleaved/simultaneous)\\
T1/T2*/T2Hahn\\
Flipping\\
Stability scan
}
\\
\hline
\end{tabular}
\end{table*}

SpinControl provides a node framework that can be used to maintain the
state of the system by performing configurable actions, such as
experiments, checking conditions, or updating calibration values. The
individual result of these actions is stored within the node's
\texttt{NodeResults}. The \texttt{NodeStatus} of that node is updated
to reflect the status of a single system state.

The basic node framework, located in the \textit{utils} package, provides
interfaces and base classes that allow full customization and
definition of new node types and action types. SpinControl has a set of
predefined nodes such as \texttt{CalibrationNode},
\texttt{CheckNode}, \texttt{BenchmarkNode}, and
\texttt{SequenceNode}. Each of these node types implements specific
logic (supported by action types) to fulfil its task and determine its
\texttt{NodeStatus}. The nodes allow definition of \textit{prenodes} to
let a node maintain other nodes before executing further actions.
Execution of those actions can then depend on the result of those
prenodes, allowing the definition of a graph-like structure if needed
\cite{Jul18}.

Nodes are configurable by the user through a configuration file. This
is where specific nodes are selected (\texttt{CalibrationNode},
\texttt{BenchmarkNode}, etc.) and their actions configured to hold e.g.
a specific experiment implementation that supports the required node
actions (e.g. a calibration node would need a `check',
`calibrate\textquotesingle, and `update' \texttt{Action}). An
\texttt{Action} can refer to a method of the
\texttt{LlsExperiment} class, e.g. to measure through the
\texttt{MeasurementManager} or analyse with an \textit{analysis} class,
or any other function.

Nodes can be implemented by the user through extension of a node base
class. A node is required to implement a \texttt{maintain()} method
that accepts the global \texttt{MaintenanceState}. This
\texttt{MaintenanceState} object holds the full node configuration as
well as all node results since the start of maintenance and allows each
node to inspect the results prior to its own \texttt{maintain()} call.

The root object that starts node maintenance is the
\texttt{NodeMaintainer}, which can simply maintain one or more nodes and
catch and handle any errors that may occur. The
\texttt{NodeMaintainer}, like any framework class, can be defined or
extended by the user.

To support custom backend behaviour, the framework also provides an
extensible \texttt{ActionExecutor}. This object should provide
system-specific functions, such as logging each action to a log file,
or handle calibration exceptions in a specific way.

SpinControl provides default implementations of the node framework
classes and supports automated calibration through its
\texttt{CalibrationManager}, mostly by using the standardized
experiments from \textit{lls-experiments}. The node framework can also be
used for automated tuning, but no nodes are currently implemented for
this. The relations between the classes are shown in
Fig.~\ref{fig7}.

\Figure[t!](topskip=0pt, botskip=0pt, midskip=0pt)[width=3.07684in]{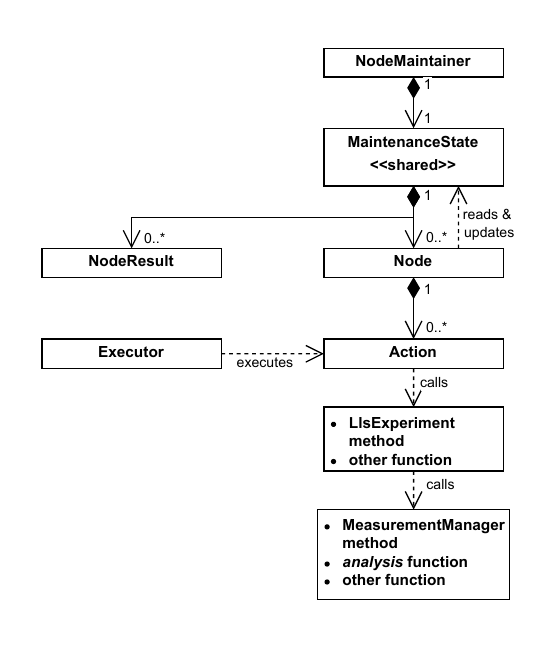}
{\label{fig7} Class diagram, showing the NodeMaintainer having a MaintenanceState with NodeResults and Nodes, the latter of which having Actions. The Actions call methods of LlsExperiment or other functions. All Nodes have access to the MaintenanceState. An Action is executed by an Executor.}

\subsection{Orchestration}\label{orchestration}

The main orchestration of all components is done by \texttt{LlsMain}.
\texttt{LlsMain} retrieves configuration and sets up the components
described earlier, including possible interdependencies. It is the main
access point to the components and contains proxies of important
settings.

The \texttt{OnlineManager} component of SpinControl handles the network
connection to a local component of the Quantum Inspire cloud platform.
This local component handles job queuing and the connection to the web
frontend or Python SDK. The \texttt{OnlineManager} sends static
information about the spin backend such as available number of qubits,
connectivity, and the primitive gate set, and dynamic information such
as coherence times. It polls for quantum algorithms in the form of
cQASM strings, passes them to the \texttt{MeasurementManager}, before
returning the measured multi-qubit bitstring counts and fractions.

The \texttt{CalibrationManager} handles auto-calibration, starting the
\texttt{NodeMaintainer} and maintaining \texttt{Cali\-bration\-Node}
objects with a set interval. \texttt{LlsMain} has a state machine
\texttt{LlsStateMachine} with two states: one for `online' and
`offline', and one for `calibrating', `executing', `locked', or
`waiting'. The \texttt{CalibrationManager} ensures no calibration is
started during the execution of a job from the cloud platform and vice
versa. Internal users can modify the state to manually perform
measurements or calibrations.

Fig.~\ref{fig8} shows the orchestration between the
\texttt{Online\-Manager}, \texttt{Calibration\-Manager}, and
\texttt{Measurement\-Manager}.

\Figure[t!](topskip=0pt, botskip=0pt, midskip=0pt)[width=7in]{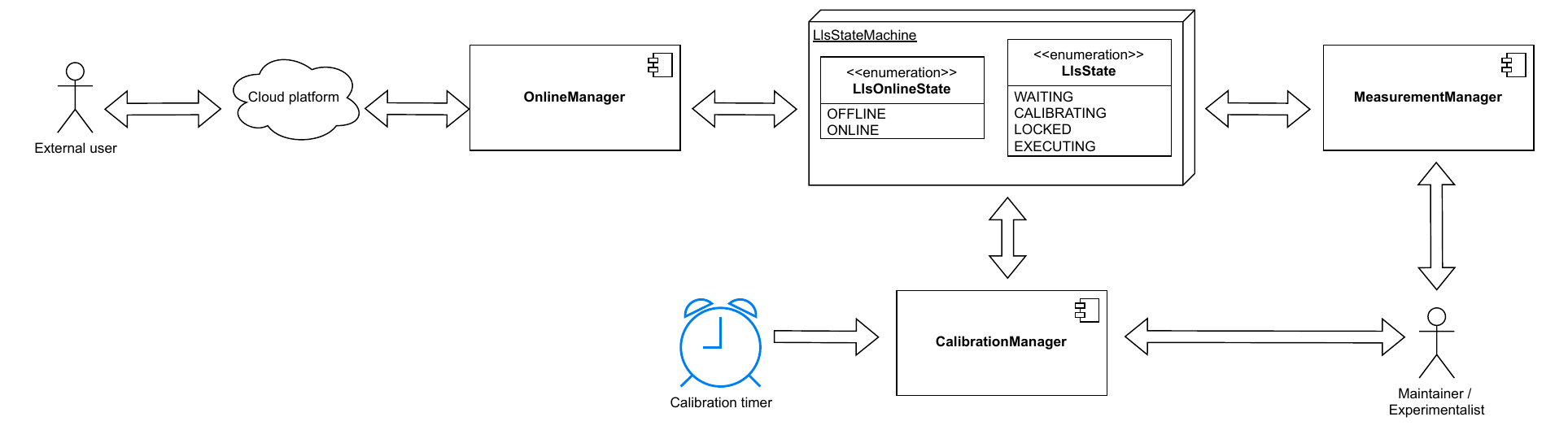}
{\label{fig8} Orchestration between: algorithms from external users through the cloud
platform and OnlineManager, periodic calibrations through the
CalibrationManager, and measurements from internal users through the
CalibrationManager and MeasurementManager, guided by the online and
operational states.}

\subsection{Inputs/outputs/GUIs}\label{inputsoutputsguis}

The internal user interacts with SpinControl through in-memory and
persistent inputs and outputs, and graphical user interfaces (GUIs).
Inputs for experiments, or for measurements and analyses separately, are
provided in the form of in-memory dataclasses.

Static configuration is provided by configuration files:

\begin{itemize}
\item
  Instrumentation:

  \begin{itemize}
  \item
    Instrument definitions: holds the instrument driver classes and
    arguments to instantiate them
  \item
    Instrument settings: holds collections of settings to be performed
    on instruments around measurements, where the
    \texttt{startup\_settings} are performed at startup of
    \texttt{LlsMain}
  \item
    Port definitions: mapping of DAC instrument channel to device gate
  \item
    VPort definitions: virtual gate matrices
  \end{itemize}
\item
  Compilation:

  \begin{itemize}
  \item
    Compiler settings: holds the mapping of AWG instrument channel to
    device gate, and settings such as delays. It also holds the mapping
    of lock-in amplifier instrument channel to readout name, and
    settings such as modulation frequency and amplitude
  \item
    Device definition: holds the available qubits and qubit connectivity
    and the \texttt{DeviceManager} plug-in to register quantum
    operations on the devices (e.g. \texttt{X} or \texttt{Read})
  \end{itemize}
\item
  Experiments and calibration:

  \begin{itemize}
  \item
    Calibration settings: holds the plug-in to build the
    \texttt{Node} structure by the \texttt{CalibrationManager}
    and settings such as the calibration period and number of retries
  \item
    Datastore definitions: holds the types and descriptions of the
    entries in the datastore
  \end{itemize}
\item
  Orchestration:

  \begin{itemize}
  \item
    Main settings: holds general settings such as the experiment output
    directory, logging configuration, and settings to connect to the
    cloud platform.
  \end{itemize}
\end{itemize}

The outputs of experiments are returned in memory and are stored on disk
in a directory structure by year, day, and experiment-unique identifier,
following the structure of the \textit{quantify} package \cite{Ora}. Each
experiment can be set to return a dataset with independent and dependent
variables, a snapshot with the values of instrument parameters at the
start of the measurement, a copy of the datastore, and metadata
concerning the measurement such as the measurement input definition. If
analysis is performed on the resulting dataset, the analysis (fit)
values are stored as well as plots.

The datastore is currently implemented as a JSON file that is backed up
with each update of a value.

Users can interact with SpinControl through various GUIs:

\begin{itemize}
\item
  \textbf{InstrumentMonitor}. From the \textit{quantify} package, shows
  the current values of all instrument parameters
\item
  \textbf{PortGUI}. From the \textit{core-tools} package \cite{Del26},
  shows and allows modification of the current values of the
  \texttt{PortDAC} instrument
\item
  \textbf{VGM GUI}. From the \textit{core-tools} package, shows and allows
  modification of the entries of the virtual gate matrices
\item
  \textbf{VideomodeGUI}. From the \textit{core-tools} package, allows
  performing repetitive one or two-dimensional scans using the AWG.
\item
  \textbf{CalibrationNodesGUI}. Shows the \texttt{Calibration\-Node} objects and
  their status.
\item
  \textbf{Dataviewer}. From the \textit{qt-dataviewer} package
  \cite{Del263}, allows browsing plots of the datasets with measurement
  output.
\item
  \textbf{Latest experiments GUI}. Gives a rolling view with plots from
  the latest experiments.
\item
  \textbf{ExperimentGUI}. Allows running experiments with modifiable
  values.
\item
  \textbf{DatastoreGUI}. Shows and allows modification of values in the
  datastore
\end{itemize}

\section{Conclusion}\label{conclusion}

SpinControl is a software package focused on the operation of spin-qubit
systems. Written in Python, it implements functionality specific for
spin qubits, such as standardized spin experiments and virtual gate
matrices. By separating data and behaviour, the software offers
customizability of device layout, instruments (of different vendors)
with parameters, pulse or circuit programs, and pre-storage processing
of results, as well as creation of standardized experiments and
calibration graphs. Its modular architecture makes it easier to extend
the software with new functionality. Robustness is achieved through
continuous integration pipelines and input validation. SpinControl can
be connected to the Quantum Inspire cloud platform to execute algorithms
remotely.

The long-term outlook for SpinControl revolves around maintaining the
flexibility of adding more spin-specific functionalities, such as
shuttling, dispersive gate readout, and automated tuning. We are making
SpinControl even more modular, to make it easier to integrate parts of
the library in other control software stacks and to integrate other
libraries into SpinControl. We plan to expand SpinControl to support
additional remote connections and more instruments.

\section*{Acknowledgment}\label{acknowledgment}

The authors would like to acknowledge Sander de Snoo and Lieven
Vandersypen for their support in providing the core‑tools, pulse‑lib,
QConstruct, and qt-dataviewer software packages.

This publication is part of the project ``Quantum Inspire - the Dutch
Quantum Computer in the Cloud'' with file number NWA.1292.19.194 of the
research programme ``Onderzoek op Routes door Consortia 2019'' and the
catalyst programme ``CAT-1: Quantum Inspire: the first European quantum
computer'' of the QDNL National Growth Fund research programme, which
are (partly) financed by the Dutch Research Council (NWO). This project
has received funding from the European Union's Horizon 2020 research and
innovation programme under grant agreement No 951852 and the European
Union's Horizon Europe research and innovation programme under grant
agreement No 101174557. This publication reflects the views only of the
author, and the Commission cannot be held responsible for any use which
may be made of the information contained therein.

\bibliographystyle{IEEEtran} 
\bibliography{references}

\EOD

\end{document}